\documentclass[12pt]{article}
\pdfoutput=1
\usepackage{geometry,enumerate,amsmath,amssymb}
\usepackage{fullpage}
\usepackage{graphicx}
\usepackage{bm}% bold math
\numberwithin{equation}{section}

\newcommand{\be}{\begin{equation}}
\newcommand{\ee}{\end{equation}}
\newcommand{\bea}{\begin{eqnarray}}
\newcommand{\eea}{\end{eqnarray}}

\renewcommand{\epsilon}{\varepsilon}

\newcommand{\bnabla}{\mbox{\boldmath $\nabla$}}
\renewcommand{\bm}{{\bf m}}
\begin{document}
\title{
%  \begin{flushright}\ \vskip -2cm {\small{\em DCPT-18/19}}\end{flushright}\vskip 1cm 
 Hopfions in chiral magnets
}
\author{
  Paul Sutcliffe\\[10pt]
 {\em \normalsize Department of Mathematical Sciences,}\\
 {\em \normalsize Durham University, Durham DH1 3LE, United Kingdom.}\\ 
{\normalsize Email:  p.m.sutcliffe@durham.ac.uk}
}
\date{July 2018}

\maketitle
\begin{abstract}
  A magnetic Hopfion is a three-dimensional topological soliton that consists of a closed loop of a twisted magnetic Skyrmion string.
  The results of numerical simulations are presented that demonstrate the existence of a stable Hopfion in a nanocylinder of a chiral magnet and an explicit analytic expression is shown to provide a reasonable approximation to the numerically computed Hopfion. A mechanism is suggested to create the Hopfion
from a target Skyrmion by introducing an interfacial perpendicular magnetic anisotropy.
\end{abstract}

\newpage
\section{Introduction}\quad
Topological solitons are smooth field configurations that impersonate particles \cite{MS}. A nanoscale example, where the topology is two-dimensional, is provided by a Skyrmion in a chiral magnet \cite{BY}. This has the potential for significant technological applications within information storage and logic devices \cite{Fert}, and is therefore the focus of intense theoretical and experimental research. The magnetic Skyrmion has a natural generalization to a topological soliton with a three-dimensional topology, known as a Hopfion, that may be interpreted as a closed loop of a twisted Skyrmion string \cite{NaV}. The current level of interest in magnetic Skyrmions provides a significant motivation to study the possibility of Hopfions in magnetic materials, which is a topic that has received little attention so far.

Hopfions were first studied in relativistic field theories \cite{Fa2,FN,GH}, where substantial numerical computations revealed a range of stable knotted and linked Hopfions \cite{BS5,HS,HS2,Su,Jen}. In the context of magnetic materials, simulations support the existence of nanoscale Hopfions in frustrated magnets \cite{Sut_frus}, but a suitable host material has yet to be identified with the required frustration.
In contrast, impressive experiments on
chiral ferromagnetic liquid crystal colloids have produced spectacular realizations of micrometer sized Hopfions \cite{AS} that are in good agreement with the results of numerical computations using the Frank free energy \cite{Frank}.

Naively, it is tempting to conclude that the results on Hopfions in chiral liquid crystals automatically transfer to the setting of chiral magnets, with a simple change of scale from the micrometer to the nanometer. This conclusion is based on the fact that the Frank energy includes terms describing the splay, twist and bend of the liquid crystal, and if these are given equal weight (known as the one-constant approximation) then the expression for the energy of the chiral liquid crystal exactly reproduces that of a chiral magnet, with an appropriate identification of constants.
However, this simple argument contains a fatal flaw. The one-constant approximation is not a good description of the regime used to obtain Hopfions. The weighting of the twist is lower than the other terms and this allows perpendicular anchoring conditions on the surface of a thin film to induce a stable uniform ground state in which to immerse the Hopfion, without the need for lateral confinement. In the case of a chiral magnet, the increased contribution of the twist destroys a uniform ground state, even in the presence of 
perpendicular anchoring conditions on the surface of a thin film. Unfortunately, this means that the Hopfion, as a localized disturbance embedded within a uniform state, does not survive the relocation from a liquid crystal host to a chiral magnetic host.

Here, the results of numerical simulations are presented that demonstrate a resolution of this problem. It is shown that a Hopfion can exist in a nanocylinder of a chiral magnet with dimensions similar to those used in recent experiments on target Skyrmions in the chiral magnet FeGe \cite{Zheng}. The nanocylinder is capped by layers that generate a strong interfacial perpendicular magnetic anisotropy on the flat ends of the cylinder and the curved side is found to provide a suitable lateral confinement. An explicit analytic expression is shown to provide a reasonable approximation to the numerically computed Hopfion. Finally, a mechanism is suggested to create the Hopfion from the already experimentally accessible target Skyrmion by turning on the interfacial perpendicular magnetic anisotropy.

\section{Chiral magnets and Hopfion topology}\quad
Within the standard formalism of micromagnetics, the direction of the local magnetization is given by a three-component unit vector ${\bf m}=(m_1,m_2,m_3),$ defined throughout the material. The energy density for a chiral magnet
in an external magnetic field ${\bf H}=(0,0,H)$
is given by
\be
   {\cal E}=A(\bnabla {\bf m})^2+D{\bf m}\cdot (\bnabla \times {\bf m})-M_s\bm\cdot {\bf H},
\label{energy}
   \ee
   where $A$ and $D$ are the exchange and Dzyaloshinskii-Moriya constants,
   with the constant $M_s$ being the saturation magnetization.
      This system has a natural length scale, given by the helical period $L=4\pi A/D$. For experiments in $\mbox{Fe}\mbox{Ge}$ nanocylinders \cite{Zhao} the helical period is $L=70\ \mbox{nm}.$ The relevant scale for the magnetic field is set by the critical field $H_c=D^2/(2AM_s)$, above which the ground state is the uniform state aligned with the magnetic field, $\bm=(0,0,1)={\bf e_3}.$
Most of this study will be concerned with the case of no external field, so $H=0$ from now on, unless explicitly stated otherwise.
   
At the surface of the material the free boundary conditions \cite{RT}
are those that
follow from the variation of the bulk energy density and are given by
\be({\bf n}\cdot\bnabla)\bm=\frac{2\pi}{L}  {\bf n}\times \bm,
\label{freebc}
\ee
where ${\bf n}$ is the unit outward normal to the surface.

The geometry of the chiral magnet sample is taken to be a cylinder with dimensions designed to be favourable to supporting a stable Hopfion. Explicitly, the height of the cylinder is set to the helical period $L$ and the diameter is equal to $3L.$ Coordinates are used in which the axis of the cylinder is aligned with the $z$-axis, so in terms of cylindrical coordinates $(\rho,\theta,z)$ the sample is given by $0\le \rho\le 3L/2$ and $-L/2\le z\le L/2.$ The height of the cylinder is motivated by attempts to suppress helical and cone phases and is familiar from studies on thin films. To enhance lateral confinement, the diameter of the cylinder is close to the smallest value that supports a Hopfion (found by performing numerical simulations with a range of diameters).

Motivated by the results on chiral ferromagnetic liquid crystal colloids in thin films \cite{AS},  perpendicular boundary conditions are favoured on the ends of the cylinder ($z=\pm L/2$) by capping these with a layer that yields a large interfacial perpendicular magnetic anisotropy, for example as generated at a magnetic metal/oxide interface \cite{DC}. This yields an additional contribution to the surface energy at these boundaries given by $-Km_3^2$, where $K$ is the uniaxial anisotropy constant at the interface. This modifies the free boundary condition (\ref{freebc}) at the ends of the cylinder to
\be
({\bf n}\cdot\bnabla)\bm=\frac{2\pi}{L}  {\bf n}\times \bm
+\frac{K}{A}m_3\bm\times({\bf e}_3\times \bm).
\label{ibc}
\ee
For simplicity, it is assumed that $K$ is large enough that the approximation
$\bm={\bf e}_3$ is reasonable at the ends of the cylinder $z=\pm L/2.$ Even in this strong boundary anisotropy regime, the fully polarized uniform ferromagnetic state $\bm = {\bf e}_3$ is unstable, and this is still true in the thin film limit where the diameter of the cylinder is taken to be large, so that the lateral boundaries that yield an edge twist can be ignored. As mentioned earlier, this is an important difference between the liquid crystal and magnetic systems. By introducing an external magnetic field near the critical value $H_c$, the ferromagnetic uniform ground state can be recovered but it turns out that a Hopfion is destroyed at magnetic fields well below this value, so this is not a fruitful approach. However, confinement within cylinders of the dimensions introduced above does allow stable Hopfions, as the results in the next section demonstrate.

Before presenting the numerical results, the topology of magnetic Skyrmions and Hopfions for smooth magnetizations $\bm\in S^2$ will be briefly reviewed, using standard results from homotopy theory \cite{Hilton}.
To reveal the two-dimensional topology of a Skyrmion, a plane through the three-dimensional material is chosen such that it contains a two-dimensional region $V_2$
in the material that is homeomorphic to a closed disc, with $\bm$ taking the same value at all points on the boundary
$\partial V_2$ of this region. The fact that $\bm$ takes the same value at all points on $\partial V_2$ allows the magnetization to be considered as a map
$\bm:\widetilde V_2\mapsto S^2$, where $\widetilde V_2$ denotes $V_2$ with all points on the boundary $\partial V_2$ identified. As $\widetilde V_2$ has the topology of $S^2$, such maps are classified by elements of the homotopy group
$\pi_2(S^2)=\mathbb{Z},$ counting the number of Skyrmions inside $V_2$. A single Skyrmion has this integer equal to one (with a suitable sign convention) and this corresponds to the fact that in the region $\widetilde V_2$ the magnetization  winds once around the sphere of directions, so that every possible direction is attained exactly once. Usually $V_2$ lies in a plane parallel to one of the surfaces of the material and is chosen so that
$\bm={\bf e}_3$ on $\partial V_2$. The point inside $V_2$ at which
$\bm=-{\bf e}_3$ is then defined to be the location of the Skyrmion. Of course, it may not be possible to find such a region $V_2$, because the magnetization may not contain a Skyrmion.

The three-dimensional topology of a Hopfion is a generalization of the two-dimensional topology of the Skyrmion just described. In this case the topology of a Hopfion is revealed by finding a three-dimensional region $V_3$ in the material that is homeomorphic to a closed ball, with $\bm$ taking the same value at all points on the boundary $\partial V_3$ of this region. Let $\widetilde V_3$ denote $V_3$ with all points on the boundary $\partial V_3$ identified, then this has the topology of $S^3$ and the magnetization can be considered as a map
$\bm:\widetilde V_3\mapsto S^2$. These maps are classified by elements of the homotopy group $\pi_3(S^2)=\mathbb{Z},$ that counts the number of Hopfions inside $V_3$ and is known as the Hopf charge. A single Hopfion has Hopf charge equal to one and this time the integer counts a linking number of magnetization field lines, rather than a winding number.
In detail, consider any fixed value $\bm^{(I)}$ of the magnetization, then generically the points in $\widetilde V_3$ where $\bm$ takes the value $\bm^{(I)}$ will form a closed curve $C^{(I)}.$ Consider any second fixed value
$\bm^{(II)}$, distinct from the first, and its associated closed curve $C^{(II)}.$ For a single Hopfion the curves $C^{(I)}$ and $C^{(II)}$ will be linked exactly once. Typically $V_3$ is chosen so that $\bm={\bf e}_3$ on $\partial V_3$ and then the closed curve inside $V_3$ where $\bm=-{\bf e}_3$ is defined to be the core of the Hopfion. This core may be interpreted as a closed loop of a twisted Skyrmion string, where the twist produces exactly one full rotation to generate the unit linking number between any two closed curves along which the magnetization remains constant \cite{NaV}.

\section{Numerical results and an analytic approximation}\quad
To numerically compute a Hopfion in a cylindrical chiral magnet the energy (\ref{energy}) is minimized using a standard gradient flow algorithm, which is similar to strongly damped Landau-Lifshitz-Gilbert evolution. Cylindrical coordinates are used in space as these more conveniently match to the geometry of the cylinder than Cartesian coordinates. The simulation lattice contains $106\times 64\times 71$ grid points in the $\rho,\theta,z$ directions and spatial derivatives are evaluated using second order finite difference approximations. For a cylinder of the chiral magnet FeGe of height $L$ and diameter $3L$, with $L=70\ \mbox{nm}$, this corresponds to a lattice spacing of $1\ \mbox{nm}$ in both the radial and $z$ directions.
Although cylindrical coordinates are employed, no assumptions are made regarding the angular dependence of the magnetization, so the simulations are fully three-dimensional and are not restricted to configurations with any given symmetry. The boundary conditions for the simulations are those described in the previous section, namely ${\bf m}={\bf e}_3$ at $z=\pm L/2$ and the free boundary condition (\ref{freebc}) at $\rho=3L/2.$

\begin{figure}[h]\begin{center}           
    \includegraphics[width=0.7\columnwidth]{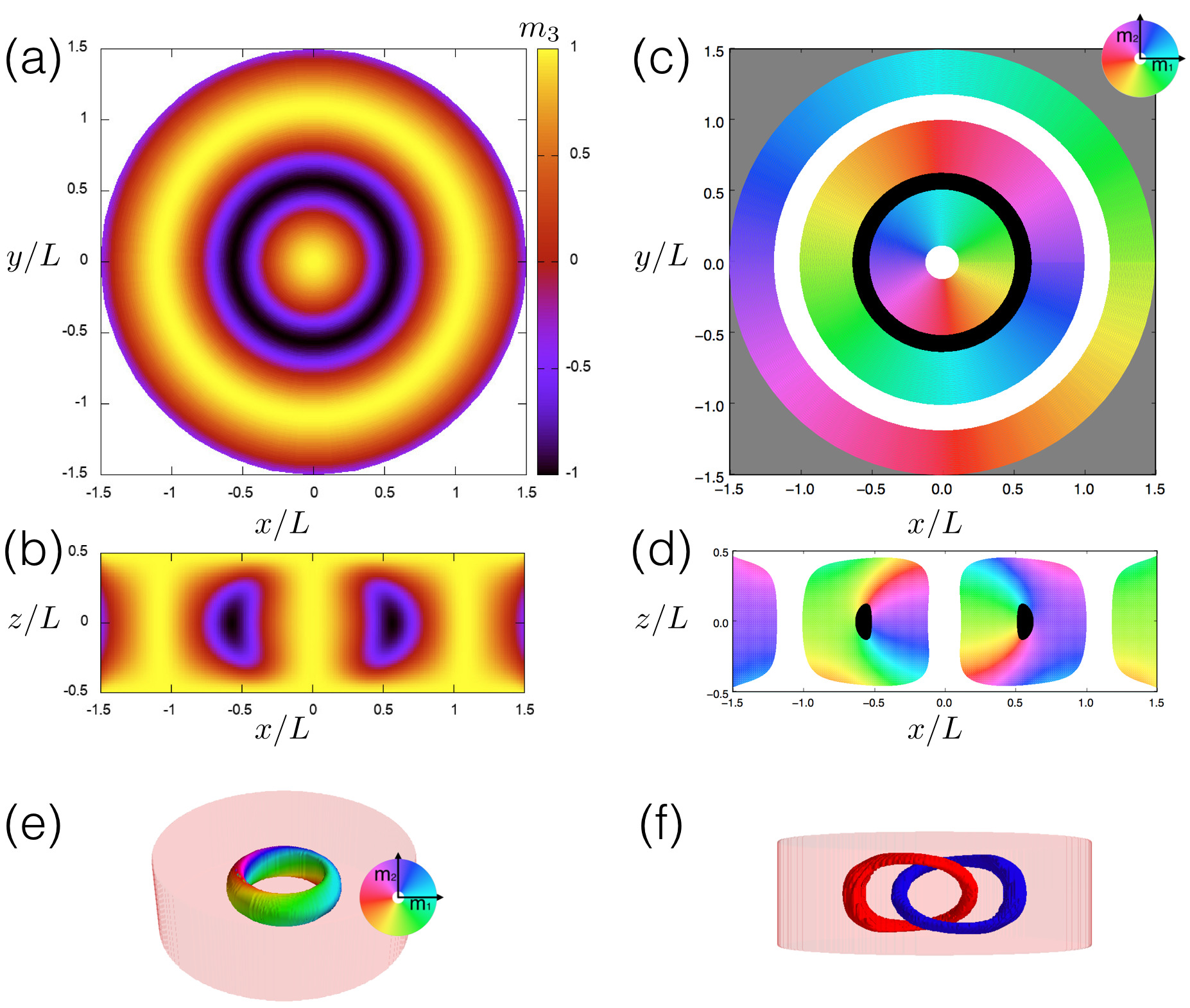}
    \caption{The numerically computed Hopfion in a cylinder of height $L$ and diameter $3L.$ 
      The magnetization component $m_3$: (a) in the plane $z=0$: (b) in the plane $y=0$. The direction of the magnetization:
      (c) in the plane $z=0$: (d) in the plane $y=0$. White denotes $m_3>0.9$ and black denotes $m_3<-0.9$. For $|m_3|\le 0.9$ the colour corresponds to the angle in the $(m_1,m_2)$-plane, as shown in the colour wheel inset. (e) shows the isosurface $m_3=-0.7$, coloured via the direction of the magnetization. (f) shows two isosurfaces $m_1=0.9$ (blue) and $m_1=-0.9$ (red).}
\label{fig1}
\end{center}\end{figure}

A variety of initial conditions yield the same final Hopfion solution after gradient flow relaxation. For example, the $z$-independent initial condition 
\be
\bm=(\sin(2\pi\rho/L)\sin\theta,-\sin(2\pi\rho/L)\cos\theta,\cos(2\pi\rho/L))
\label{ic}
\ee
can be used in the interior of the cylinder. The initial condition is not required to satisfy the boundary conditions because these are imposed directly on the cylinder surfaces. Any initial jump in the magnetization at the boundary is not an issue for the numerical simulations because gradient flow rapidly reduces this boundary mismatch.

The numerically computed Hopfion is displayed in Fig.~\ref{fig1}, where $m_3$ is shown in the plane $z=0$ in Fig.~\ref{fig1}(a) and in the plane $y=0$ in 
Fig.~\ref{fig1}(b). The black region indicates the core of the Hopfion, which may be interpreted as a circular loop of Skyrmion string. The direction of the magnetization is shown in the plane $z=0$ in Fig.~\ref{fig1}(c) and in the plane $y=0$ in Fig.~\ref{fig1}(d).  For $|m_3|\le 0.9$
the colour represents the angle in the $(m_1,m_2)$-plane according to the colour wheel inset. Regions where $m_3>0.9$ are represented by white and those where  $m_3<-0.9$ by black.

These images confirm the interpretation of the Hopfion core as a twisted Skyrmion string. In particular, the cross-section through the plane $y=0$ clearly shows the characteristic winding of a Skyrmion in the region $x>0$, as the Skyrmion string moves into the plane. The Skyrmion string returns to this plane in the region $x<0$, but this time moving out of the plane so that it appears as an anti-Skyrmion in this cross-section. The isosurface where $m_3=-0.7$ is presented in Fig.~\ref{fig1}(e), producing a torus that surrounds the core of the Hopfion. This isosurface is coloured using the same colour wheel to indicate the direction of the magnetization in the $(m_1,m_2)$-plane. A fixed colour on this torus therefore corresponds to a closed loop along which the magnetization remains constant. The fact that each colour winds once around the torus confirms that this is a Hopfion with unit Hopf charge.
The isosurface where $m_1=0.9$ is shown in blue in Fig.~\ref{fig1}(f), together with the isosurface where $m_1=-0.9$ shown in red. These two tubes surround the curves where $\bm=(1,0,0)$ and $\bm=(-1,0,0)$ respectively, and are clearly linked once, as required by the unit Hopf charge.

\begin{figure}[h]\begin{center}           
    \includegraphics[width=0.7\columnwidth]{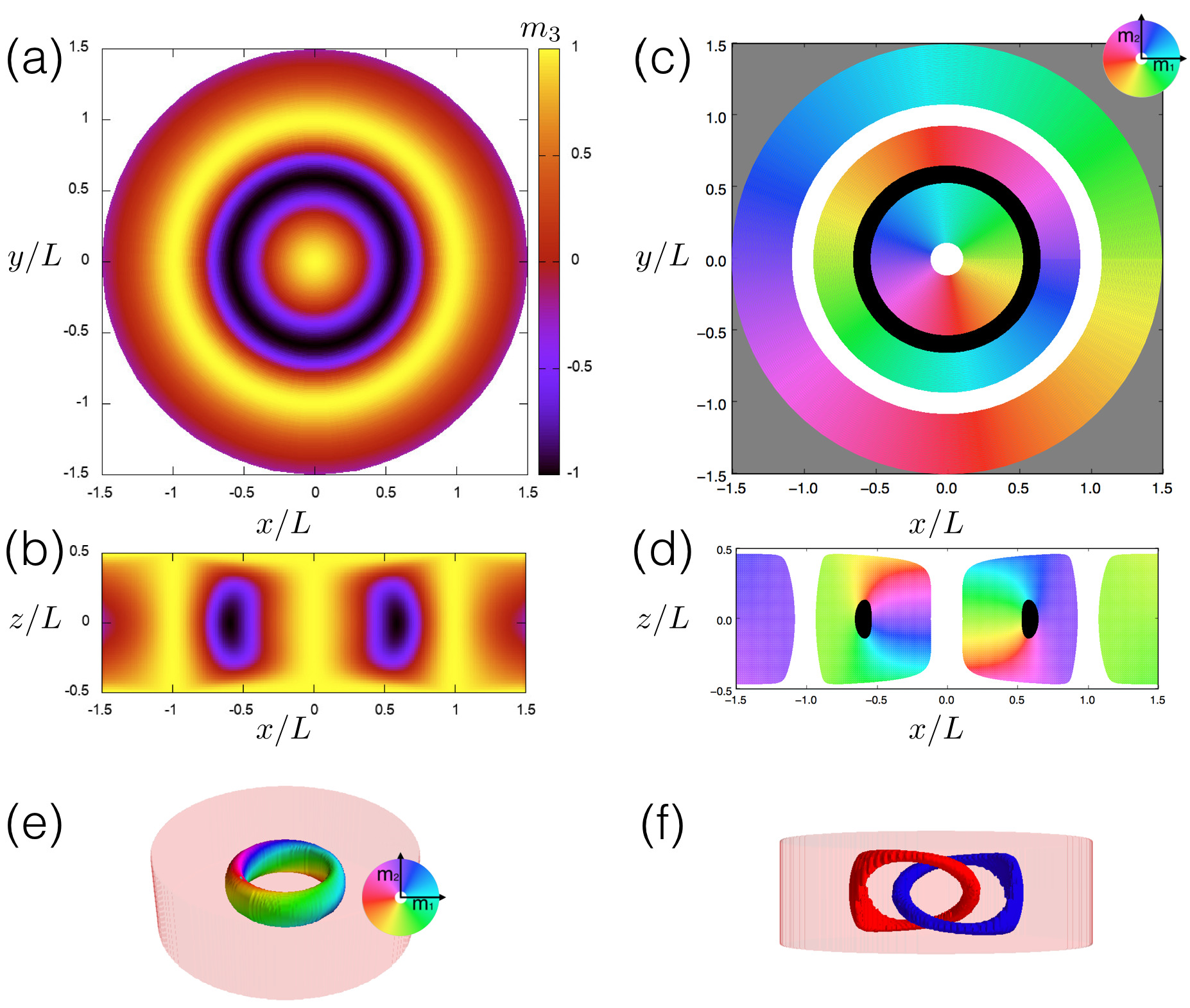}
    \caption{The same plots as in Fig.~1, but using the analytic approximation (\ref{approx}), rather than the numerically computed Hopfion.}
\label{fig2}
\end{center}\end{figure}
There is an explicit analytic expression that provides a reasonable approximation to the numerically computed Hopfion. In detail, define the three functions
$\Omega,\Xi,\Lambda$ to be
\be
\Omega=\tan(\pi z/L),\qquad
\Xi={(1+(2z/L)^2)}\sec\big(\pi\rho/(2L)\big)/L,\qquad
\Lambda =\Xi^2\rho^2+\Omega^2/4,
\ee
then the approximation is given by
\be
{\bf m}=\bigg(\frac{4\Xi\rho\big(\Omega\cos\theta-(\Lambda-1)\sin\theta\big)}{(1+\Lambda)^2},\ 
\frac{4\Xi\rho\big(\Omega\sin\theta+(\Lambda-1)\cos\theta\big)}{(1+\Lambda)^2},\
1-\frac{8\Xi^2\rho^2}{(1+\Lambda)^2}\bigg).
\label{approx}
\ee
In Fig.~\ref{fig2} the same quantities are plotted as in Fig.~\ref{fig1}, but using the above analytic approximation for the Hopfion, rather than the numerically computed Hopfion. A comparison of these two figures confirms that the approximation indeed provides a reasonable description of the Hopfion and certainly captures the main qualitative features. This approximation was obtained by taking the standard Hopf map in compactified $\mathbb{R}^3$ and performing mappings on the coordinates to better represent the geometry of the cylinder. In particular, by enforcing the boundary condition $\bm={\bf e}_3$ at $z=\pm L/2.$

If this approximation is used as an initial condition for the numerical simulation then it quickly relaxes to the Hopfion, as expected given the reasonable agreement between the two. The explicit expression (\ref{approx}) allows some properties to be studied in detail, that are expected to be reflected in the numerical solution. For example, the Hopfion is not axially symmetric, as seen by the fact that (\ref{approx}) has an explicit $\theta$ dependence. However, it is equivariant because a spatial rotation $\theta\mapsto\theta+\alpha$ induces a rotation in the $(m_1,m_2)$-plane through the same angle $(m_1+im_2)\mapsto(m_1+im_2)e^{i\alpha}$. The energy density (\ref{energy}) is not invariant under either of these rotations alone, but is invariant if both are applied. The equivariance property manifest in the approximation therefore implies that the energy density is axially symmetric. The topology is also manifest in the expression (\ref{approx}). Let $V_3$ be the cylinder given by $0\le\rho\le L$ and $|z|\le L/2.$ It is easy to check that on the boundary $\partial V_3$ of this cylinder the approximation satisfies the property that $\bm={\bf e}_3$, and hence there is an integer-valued Hopf charge for the map $\bm:\widetilde V_3\mapsto S^2$, which can be proved to be equal to one. In other words, it is precise to say that there is exactly one Hopfion in the cylinder $V_3$, even though the Hopf invariant is not defined for the magnetization extended to the larger cylinder of the material.

The images in Fig.~\ref{fig1}(a) and Fig.~\ref{fig1}(c) reveal that in the plane $z=0$ the Hopfion closely resembles a target Skyrmion \cite{Du,LRM}, of the type recently imaged directly in FeGe nanocylinders \cite{Zheng}. This suggests a mechanism to create a Hopfion from the already experimentally accessible target Skyrmion by turning on an interfacial perpendicular magnetic anisotropy. The perpendicular magnetic anisotropy at an interface between magnetic metals and oxides can be tuned by applying a voltage across the oxide layer \cite{Maetal}.

\begin{figure}[h]\begin{center}           
    \includegraphics[width=0.7\columnwidth]{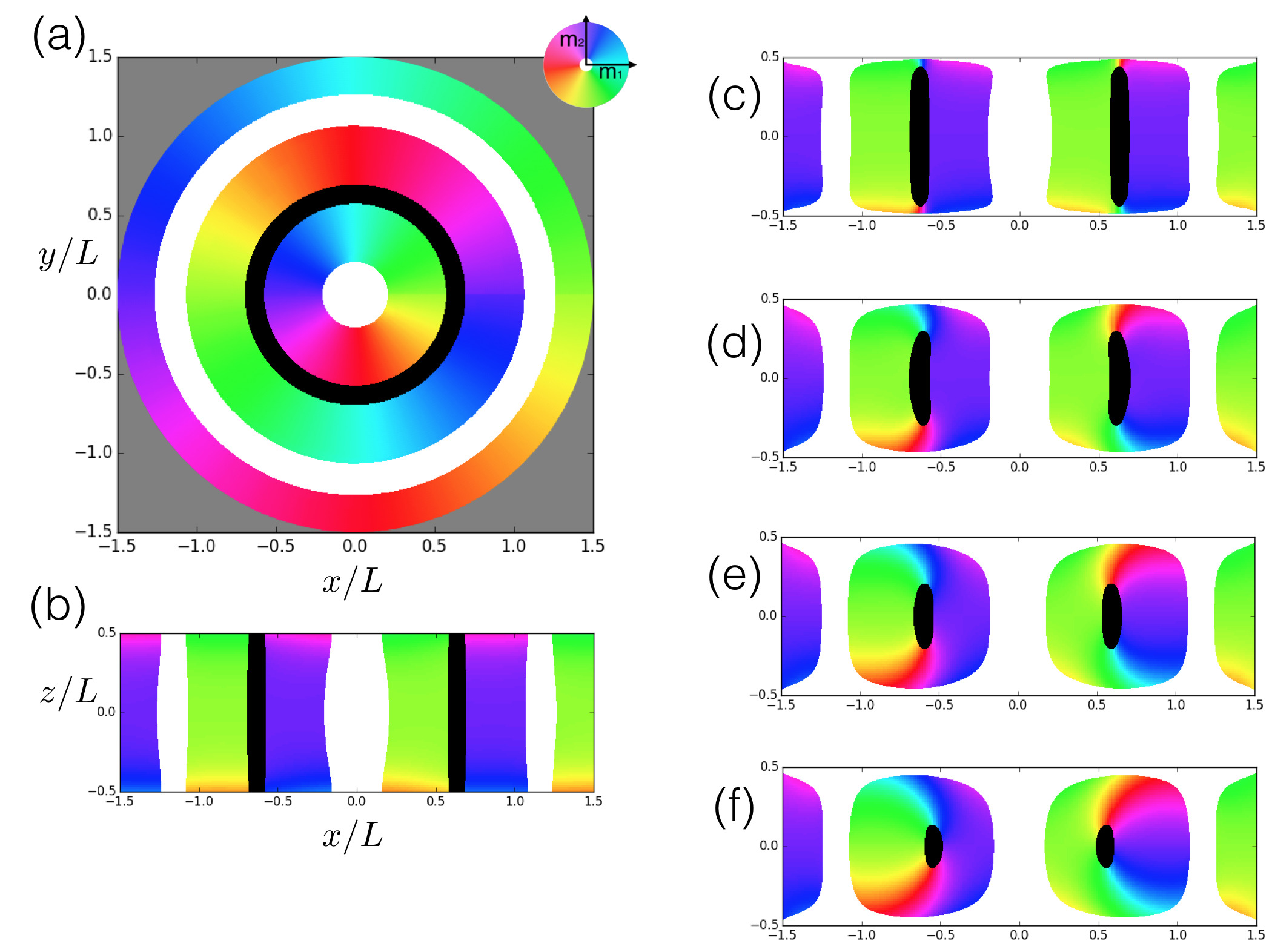}
    \caption{The transformation of a target Skyrmion into a Hopfion. The direction of the magnetization for a target Skyrmion:
      (a) in the plane $z=0$: (b) in the plane $y=0$. The colour scheme is as in Fig.~1. The direction of the magnetization in the plane $y=0$ (c,d,e,f) during gradient flow energy relaxation of the target Skyrmion into the Hopfion upon the introduction of a perpendicular magnetic anisotropy at the surfaces $z=\pm L/2.$}
\label{fig3}
\end{center}\end{figure}

The transformation of a target Skyrmion into a Hopfion is demonstrated in the simulation presented in Fig.~\ref{fig3}. To obtain the target Skyrmion the perpendicular magnetic anisotropy at the surfaces $z=\pm L/2$ is removed and the boundary condition $\bm={\bf e}_3$ at these surfaces is replaced by the free boundary condition (\ref{freebc}). An external magnetic field $H=H_c/4$ is also introduced to suppress a helical state \cite{Letal}. However, this is not required if the
energy (\ref{energy}) also includes the contribution from the demagnetization energy \cite{Zheng}.

The direction of the magnetization for the target Skyrmion is shown in the plane $z=0$ in Fig.~\ref{fig3}(a) and in the plane $y=0$ in Fig.~\ref{fig3}(b). Note that the magnetization for the target Skyrmion is not quite independent of $z$ because of the chiral surface twists \cite{Letal} that appear close to the boundaries $z=\pm L/2$ and are evident in the colour changes near these boundaries.
Fig.~\ref{fig3}(c,d,e,f) shows the evolution of the magnetization in the plane $y=0$ under gradient flow energy relaxation starting from the target Skyrmion once the external magnetic field is removed and the perpendicular magnetic anisotropy is introduced by enforcing the boundary condition  $\bm={\bf e}_3$ at $z=\pm L/2$. The initial cylinder along which ${\bf m}=-{\bf e}_3$ retreats to a circle and the magnetization twists around this circle to form the twisted Skyrmion string comprising the Hopfion. There is very little change in the plane $z=0$, which is why the Hopfion resembles a target Skyrmion in this plane. 

\section{Conclusion}\quad
Numerical simulations support the existence of a stable Hopfion in a nanocylinder of a chiral magnet such as FeGe in the presence of an interfacial perpendicular magnetic anisotropy at the ends of the nanocylinder. Target Skyrmions have already been obtained experimentally in similar nanocylinders without the interfacial perpendicular magnetic anisotropy and computations predict that introducing the anisotropy will transform the target Skyrmion into a Hopfion, thus providing a viable experimental procedure for Hopfion creation. Given the potential technological applications of magnetic Skyrmions it seems likely that there may be a role to play for their Hopfion relatives that could exploit the fully three-dimensional nature of these topological solitons. Future studies are required to fully investigate the properties of Hopfions in chiral magnets, for example by considering their dynamical properties under varying magnetic fields and their response to electric currents.

For simplicity, it has been assumed that at the ends of the cylinder
the uniaxial anisotropy constant at the interface
is large enough that the approximation $\bm = {\bf e}_3$ is reasonable. 
However, this approximation is not necessary and applying the boundary condition (\ref{ibc}) directly yields similar results if $K$ is sufficiently large, of the order of $D$, which is realistic for physical material parameters. Finally, the demagnetization energy has been neglected in this study but the expectation is that this should only improve the stability of the Hopfion, in a similar way to the situation for target Skyrmions \cite{Zheng}.

\

\noindent{\em Note added:} During the completion of this work two preprints 
\cite{TS,LLZ} appeared on the arXiv with a substantial overlap with the contents of this paper. There is a good agreement between the results in all three papers even though different methods are employed in each. 

\section*{Acknowledgements}
\noindent
Many thanks to Yizhou Liu, Ivan Smalyukh and Jiadong Zang for useful discussions. This work is funded by the
Leverhulme Trust Research Programme Grant RP2013-K-009, SPOCK: Scientific Properties Of Complex Knots.


\begin{thebibliography}{99}

\bibitem{MS} N.S. Manton and P.M. Sutcliffe, \textit{Topological Solitons}, CUP (2004).

\bibitem{BY}
 A.N. Bogdanov  and D.A Yablonskii,
%  Thermodynamically stable vortices in magnetically ordered crystals: The mixed state of magnets.
  \textit{Sov. Phys. JETP} {\bf 68}, 101 (1989).  

  \bibitem{Fert}
  A. Fert, V. Cros and J. Sampaio,
%  Skyrmions on the track.
  \textit{Nat. Nanotechnology} {\bf 8}, 152 (2013).

\bibitem{NaV}
 P.M. Sutcliffe,
 \textit{Nat. Mater.} {\bf 16}, 392 (2017).
  
\bibitem{Fa2} L.~D. Faddeev,
{\em Quantization of solitons},
Princeton preprint IAS-75-QS70 (1975).

\bibitem{FN} L. Faddeev and A.~J. Niemi,
%Stable knot-like structures in classical field theory,
    \textit{Nature} \textbf{387}, 58 (1997).

\bibitem{GH} J. Gladikowski and M. Hellmund,
%Static solitons with nonzero Hopf number,
  \textit{Phys. Rev.} \textbf{D56}, 5194 (1997).

\bibitem{BS5} R.~A. Battye and P.~M. Sutcliffe,
%Knots as stable soliton solutions in a three-dimensional classical field theory,
\textit{Phys. Rev. Lett.} \textbf{81}, 4798 (1998).

  \bibitem{HS} J. Hietarinta and P. Salo,
%Faddeev-Hopf knots: dynamics of linked un-knots,
\textit{Phys. Lett.} \textbf{B451}, 60 (1999).

\bibitem{HS2} J. Hietarinta and P. Salo,
%Ground state in the Faddeev-Skyrme model,
  \textit{Phys. Rev.} \textbf{D62}, 081701(R) (2000).

\bibitem{Su} 
 P.~M. Sutcliffe,
%  Knots in the Skyrme-Faddeev model.
 \textit{Proc. R. Soc.} {\bf A463}, 3001 (2007).

\bibitem{Jen} P. Jennings,
  \textit{J. Phys.} \textbf{A48}, 315401 (2015).

\bibitem{Sut_frus} P.~M. Sutcliffe,
  \textit{Phys. Rev. Lett.} \textbf{118}, 247203 (2017).

\bibitem{AS}
  P.J. Ackerman and I.I. Smalyukh,
  \textit{Nat. Mater.} {\bf 16}, 426 (2017).

\bibitem{Frank}
  F.C. Frank,
  \textit{Discuss. Faraday Soc.} {\bf 25}, 19 (1958).
    
  \bibitem{Zheng}
  F. Zheng, et al.,
  % Direct Imaging of a Zero-Field Target Skyrmion and Its Polarity Switch in a Chiral Magnetic Nanodisk,
  \textit{Phys. Rev. Lett.} {\bf 119}, 197205 (2017).

  \bibitem{Zhao}
X. Zhao, et al.,
  %Direct imaging of magnetic field-driven transitions of skyrmion cluster states in FeGe nanodisks, 
  \textit{Proc. Natl. Acad. Sci.} {\bf 113}, 4918 (2016).

\bibitem{RT}
  S. Rohart and A. Thiaville,
 % Skyrmion confinement in ultrathin film nanostructures in the presence of Dzyaloshinskii-Moriya interaction
\textit{Phys. Rev.} {\bf B88}, 184422 (2013).
  
  \bibitem{DC}
    B. Dieny and M. Chshiev,
 %   Perpendicular magnetic anisotropy at transition metal/oxide interfaces and applications
    \textit{Rev. Mod. Phys.} {\bf 89}, 025008 (2017).

  \bibitem{Hilton}
    P.J. Hilton,
    \textit{An Introduction to Homotopy Theory}, CUP (1953).

\bibitem{Du}
  H. Du, W. Ning, M. Tian and Y. Zhang,
 % Magnetic Vortex with Skyrmionic Core in Thin Nanodisk of the Chiral Magnets
  \textit{EPL} {\bf 101}, 37001 (2013).

\bibitem{LRM}
  A.O. Leonov, U.K. R\"o{\ss}ler and M. Mostovoy,
%  Target-skyrmions and skyrmion clusters in nanowires of chiral magnets
  \textit{EPJ Web Conf.} {\bf 75}, 05002 (2014).


  \bibitem{Maetal}
 T. Maruyama, et al., 
% Large voltage-induced magnetic anisotropy change in a few atomic layers of iron
 \textit{Nat. Nanotech.} {\bf 4}, 158 (2009).

 \bibitem{Letal}
  A.O. Leonov, et al., 
  %Chiral Surface Twists and Skyrmion Stability in Nanolayers of Cubic Helimagnets
  \textit{Phys. Rev. Lett.} {\bf 117}, 087202 (2016).

 
\bibitem{TS}
 J-S.B. Tai and I.I. Smalyukh, arXiv:1806.00453. 
  
\bibitem{LLZ}
 Y. Liu, R. Lake and J. Zang, arXiv:1806.01682.

  
  \end{thebibliography}
\end{document}